\documentclass[twocolumn]{aastex631}

\usepackage{multirow}
\usepackage{wasysym}
\usepackage{enumerate}
\usepackage{amssymb} 
\shorttitle{MIRI Detects Light from WD~1856\,b}
\shortauthors{Limbach et al.}
\graphicspath{{./}{figures/}}

\begin{document}
\title{Thermal Emission and Confirmation of the Frigid White Dwarf Exoplanet WD 1856+534~b}
\correspondingauthor{Mary Anne Limbach}
\email{mlimbach@umich.edu}

\author[0000-0002-9521-9798]{Mary Anne Limbach}
\affiliation{Department of Astronomy, University of Michigan, Ann Arbor, MI 48109, USA}

\author[0000-0001-7246-5438]{Andrew Vanderburg}
\affiliation{Department of Physics and Kavli Institute for Astrophysics and Space Research, Massachusetts Institute of Technology, Cambridge, MA 02139, USA}

\author[0000-0003-4816-3469]{Ryan J. MacDonald}
\altaffiliation{NHFP Sagan Fellow}
\affiliation{Department of Astronomy, University of Michigan, Ann Arbor, MI 48109, USA}

\author[0000-0002-7352-7941]{Kevin B. Stevenson}
\affiliation{Johns Hopkins APL, 11100 Johns Hopkins Rd, Laurel, MD 20723, USA}

\author[0000-0001-9827-1463]{Sydney Jenkins}
\affiliation{Department of Physics and Kavli Institute for Astrophysics and Space Research, Massachusetts Institute of Technology, Cambridge, MA 02139, USA}

\author[0000-0002-9632-1436]{Simon Blouin}
\affiliation{Department of Physics and Astronomy, University of Victoria, Victoria, BC V8W 2Y2, Canada}

\author[0000-0003-3963-9672]{Emily Rauscher}
\affiliation{Department of Astronomy, University of Michigan, Ann Arbor, MI 48109, USA}

\author[0000-0001-5831-9530]{Rachel Bowens-Rubin}
\affiliation{Department of Astronomy, University of Michigan, Ann Arbor, MI 48109, USA}

\author[0000-0001-5802-6041]{Elena Gallo}
\affiliation{Department of Astronomy, University of Michigan, Ann Arbor, MI 48109, USA}

\author[0000-0001-5864-9599]{James Mang}
\altaffiliation{NSF Graduate Research Fellow.}
\affiliation{Department of Astronomy, University of Texas at Austin, Austin, TX, USA}

\author[0000-0002-4404-0456]{Caroline V. Morley }
\affiliation{Department of Astronomy, University of Texas at Austin, Austin, TX, USA}

\author[0000-0001-6050-7645]{David K. Sing}
\affiliation{Department of Earth \& Planetary Sciences, Johns Hopkins University, Baltimore, MD, USA}
\affiliation{William H.\ Miller III Department of Physics \& Astronomy, Johns Hopkins University, 3400 N Charles St, Baltimore, MD, USA}

\author[0000-0003-3987-3776]{Christopher O'Connor}
\affiliation{Center for Interdisciplinary Research and Exploration in Astrophysics (CIERA), Northwestern University, 1800 Sherman Ave., Evanston, IL 60201, USA}

\author[0000-0002-8400-1646]{Alexander Venner}
\affiliation{Centre for Astrophysics, University of Southern Queensland, Toowoomba, QLD 4350, Australia}

\author[0000-0002-8808-4282]{Siyi Xu}
\affil{Gemini Observatory/NSF's NOIRLab, 670 N. A'ohoku Place, Hilo, Hawaii, 96720, USA}

\begin{abstract}
We report the detection of thermal emission from and confirm the planetary nature of WD 1856+534~b, the first transiting planet known to orbit a white dwarf star. Observations with JWST’s Mid-Infrared Instrument (MIRI) reveal excess mid-infrared emission from the white dwarf, consistent with a closely-orbiting Jupiter-sized planet with a temperature of $186^{+6}_{-7}$ K. We attribute this excess flux to the known {giant} planet in the system, making it the coldest exoplanet from which light has ever been directly observed. These measurements constrain the planet’s mass to no more than six times that of Jupiter, confirming its planetary nature and ruling out previously unexcluded low-mass brown dwarf scenarios.
WD~1856+534\,b is now the first intact exoplanet confirmed within a white dwarf’s “forbidden zone,” a region where planets would have been engulfed during the star’s red giant phase. Its presence provides direct evidence that planetary migration into close orbits—including the habitable zone—around white dwarfs is possible. With an age nearly twice that of the Solar System and a temperature akin to our own gas giants, WD~1856+534\,b demonstrates JWST’s unprecedented ability to detect and characterize cold, mature exoplanets, opening new possibilities for imaging and characterizing these worlds in the solar neighborhood.

\end{abstract}

\keywords{Infrared excess, Extrasolar gas giant planets, White dwarfs}

\section{Introduction} \label{sec:intro}

Detecting light from exoplanets, whether through reflection \citep{2009Natur.459..543S,2009Sci...325..709B,2009A&A...501L..23A} or thermal emission \citep[e.g.,][]{2005Natur.434..740D,2005ApJ...626..523C, Lagrange2010,2010Natur.468.1080M}, is essential for constraining their physical characteristics. By measuring an exoplanet’s brightness and analyzing spectral absorption or emission features, we can infer its formation history \citep{2022ApJ...934...74M}, atmospheric properties \citep{2007Natur.447..183K}, and chemical composition \citep{2007Natur.445..892R,2023Natur.614..659R}. For terrestrial exoplanets, characterizing their reflected and emitted light may one day enable the search for biosignatures beyond the Solar System \citep{2020arXiv200106683G,2022A&A...664A..21Q}. However, directly detecting exoplanetary light remains challenging due to the overwhelming brightness of their host stars. Consequently, detections have primarily been achieved for hot, self-luminous planets, where the contrast ratio is more favorable—either through phase curve or eclipse monitoring for close-in planets or via direct imaging for those on wider orbits. To date, emission from no planets cooler than 275\,K—a temperature comparable to Earth—have been directly detected \citep{2024Natur.633..789M}.

White dwarf (WD) systems present a unique opportunity for the detection and characterization of cold planets. The low luminosity of WDs significantly reduces the contrast challenges that typically hinder direct detections around their main-sequence counterparts. As the evolutionary remnants of stars like the Sun, WDs offer insight into the fate of planetary systems after stellar death. Understanding how planets interact with and survive post-main-sequence evolution provides crucial information on orbital stability, dynamical migration, and potential planetary engulfment. Discovering and characterizing exoplanets in these systems can inform demographic studies of planets around evolved stars \citep{2005ApJ...633.1168D,veras2021planetary,2023A&A...675A.184L,2024AJ....167..257P} and help determine whether planets can persist through stellar evolution or if they are disrupted during the red giant phase \citep{2002ApJ...572..556D,2010ApJ...722..725Z}. Moreover, investigating WD planetary systems could provide insight into whether habitable conditions can exist around stellar remnants \citep{2011ApJ...731L..31A,2012ApJ...757L..15F,2013AsBio..13..279B,2013MNRAS.432L..11L,2018ApJ...862...69K,2020ApJ...894L...6K,2023ApJ...945L..24B,2024arXiv240603189Z}.

Little is known about the survival and migration of planets around WDs, and the survivability of exoplanets through post-main-sequence evolution depends on their initial orbital distance. Planets beyond $\sim$2\,AU are expected to remain intact \citep{Nordhaus2013MNRAS}, and a handful of planets have been detected at these wider separations \citep{2011ApJ...730L...9L,1993ApJ...412L..33T,2003Sci...301..193S,2021Natur.598..272B,2024ApJ...962L..32M,2024NatAs...8.1575Z}.
Discoveries of planets within the 2\,AU threshold provide valuable insights into late-stage planetary migration and interactions. Evidence of an evaporating giant planet producing a volatile-rich gaseous debris disk \citep{2019Natur.576...61G} around WD~J091405.30+191412.25 suggests that some massive companions may migrate inward toward close-in orbits around WDs.  \citet{Vanderburg_2020} also identified a very low-mass companion transiting WD~1856+534 at just 0.02\,AU. However, neither planet candidate around WD~J091405.30+191412.25 nor WD~1856+534 has been confirmed with additional methods. In the case of WD~1856+534, they were only able to place an upper limit on its thermal emission at the time, leaving its classification uncertain between a giant planet and a low-mass brown dwarf.

In this paper, we confirm WD~1856+534\,b using the Infrared (IR) excess method with data from JWST Mid-Infrared Instrument (MIRI). White dwarfs are similar to Earth in size --- much smaller than gas giant planets. Consequently, WD~1856\,b is expected to have mid-infrared brightness comparable to the white dwarf, despite its significantly cooler predicted temperature \citep{Limbach22}.
IR excess observations have been instrumental in identifying cold companions to WDs, as the near-featureless IR spectra of WDs make them particularly suitable for detecting faint, substellar objects within their systems. This technique has led to the discovery of various WD companions \citep{2005ApJ...632L.115K,2005ApJ...632L.119B,2008ApJ...674..431F,2012ApJ...760...26B,2020ApJ...898L..35S,Limbach22}. The first brown dwarf-WD system was uncovered via IR excess \citep{1987Natur.330..138Z,2022Natur.602..219C}, and more recently, this approach has been used to identify potential exoplanets around nearby WDs \citep{2024ApJ...973L..11L}.

In this work, we utilize the unprecedented sensitivity of JWST to detect thermal emission from and confirm the planetary nature of WD~1856+534\,b. These observations enable us to measure the planet's temperature and thereby constrain its mass, representing the first definitive detection of light from a mature planet colder than 200\,K. This detection also provides the first direct confirmation that planets can migrate into and remain intact in close orbits near the habitable zone of WDs. 
The structure of this paper is as follows: Section~\ref{sec:obs} details the JWST observations, Section~\ref{DataRed} describes the data reduction and analysis, Section~\ref{sec:results} discusses our results, and Section~\ref{sec:discuss} details our conclusions.

\section{Observations} \label{sec:obs}

The observations presented here are from the first epoch of a two-epoch Cycle 3 JWST program, GO \#5204 \citep{2024jwst.prop.5204L}. This program has two main objectives: (1) to measure the emission from the exoplanet WD~1856+534\,b (as presented in this paper), and (2) to search for additional spatially resolved companions to the white dwarf via common proper motion, which requires the second epoch of observations scheduled for July 2025. 

The first epoch of data was collected on July 29, 2024, and we observed WD~1856+534 (hereafter WD~1856) using the JWST MIRI instrument \citep{2024jwst.prop.5204L}. This included MIRI imaging in the following seven filters: F560W, F770W, F1000W, F1130W, F1280W, F1500W, and F1800W, with central wavelengths $\lambda_0$ = 5.6, 7.7, 10.0, 11.3, 12.8, 15.0 and 18.0\,µm. For all imaging, we used the fast readout mode and a four point cycling dither (starting point 1 { and going to point 4}). We chose a 4-point cycling dither to correct for bad pixels and remove background noise. We used the full MIRI imaging array, which has a 74" $\times$ 113" useable FOV, and a detector plate scale of 0.11"/pixel \citep{2015PASP..127..612B}.  The integration { times were: 1.8\,min at F560W and F770W, 2.2\,min at F1000W, 6.7\,min at F1130W, 4.6\,min at F1280W, 7.4\,min at F1500W and 22.6\,min at F1800W}, with the integration times generally increasing at the longer wavelengths to achieve SNR $\approx$ 50 with both epochs of data combined. The total integration time on the system was 47.2\,minutes. For our observations, we used between 10 and 40 groups per integration. The white dwarf is relatively faint (G=16.9\,mag), and it remained well below saturation in all observations, generally occupying less than 10\% of the detector's well depth. At the longest wavelengths, the sky background emission was substantially brighter than the white dwarf.

The timing of the observation was constrained to avoid transit or eclipse events, but the exact phase of the observation was not restricted further. The data were obtained at an orbital phase of $\phi = 0.32-0.35$, slightly past quarter phase, such that more of the planet's dayside was visible than the nightside. A false-color image is shown in { Appendix \ref{AppB}}. The image captures a physical extent of 1835\,AU $\times$ 2802\,AU for the system, which is situated at a distance of 24.8\,pc.
This system includes a white dwarf star hosting a (spatially unresolved) transiting gas giant exoplanet, WD~1856\,b, located at 0.02\,AU from the star \citep{Vanderburg_2020}, along with two M-dwarf stars (G~229-20) positioned at a much wider projected separation of 1030\,AU from the white dwarf.

\begin{table*}
\centering
\caption{ Summary of parameters for the WD~1856 system.} \label{parameters}
\medskip
\begin{tabular}{lccc}
\hline
Parameter & Value & Unit & Source \\
\hline
\textbf{White Dwarf Stellar Properties} &  &  \\
Spectral Type & DA & -- & \cite{2021AJ....162..296X}\\
Mass & 0.605$\pm$ 0.013 & M$_\odot$ & this work\\
Radius & 0.0121$\pm$ 0.0002 & R$_\odot$  & this work\\
Surface Gravity & 8.05 $\pm$ 0.02 & $\log{g_{\rm cgs}}$ & this work\\
Effective Temperature & 4920 $\pm$ 50 & K & this work\\
Cooling Age & 5.4$\pm$0.7  & Gyr & this work\\
Total System Age & $7.4-10$ & Gyr & this work\\
\textbf{Planet Properties}  & &  \\
Semimajor axis & 0.02085$\pm$0.0014 & AU &  this work$^{a}$\\
Orbital Period & 1.407939217$\pm$0.000000016 & days & \cite{2023MNRAS.521.4679K}\\
$t_0$ (minimal covariance) & 2459038.4358981$\pm$0.00000114 & BJD\_TDB & \cite{2023MNRAS.521.4679K}\\
Radius ratio (R$_{\rm p}$/R$_{*}$) & 7.86$\pm$0.01 & -- & \cite{2021AJ....162..296X} \\
Planet Radius  & 0.946$\pm$0.017 & R$_{\rm Jup}$ &  this work \\
Mid-IR Brightness Temperature$^{b}$ & 186$^{+6}_{-7}$ & K &  this work\\
Effective Temperature$^{b}$  & 184$\pm$8 & K &  this work\\
Equilibrium Temperature$^{c}$ & 165$\pm$9 & K & this work\\
Mass$^{d}$ & 5.2$^{+0.7}_{-0.8}$ & M$_{\rm Jup}$ & this work\\
\hline
\end{tabular}     \\ 
      $^{a}$Computed with Kepler's third law.\\
      $^{b}$The mid-IR brightness temperature is from the blackbody fit, while the effective temperature is calculated based on the atmospheric retrieval.\\
      $^{c}$Equilibrium temperature is given by $T_{eq}=T_{\rm WD,eff}(1 - \alpha)^{1/4}\sqrt{\frac{R_\star}{2a}}$ assuming an albedo of $\alpha$ = 0.3 similar to Jupiter\citep{1981JGR....86.8705H}.\\
      $^{d}$Upper limit; assumes insolation contributes negligibly to its thermal evolution, with temperature derived from evolutionary models based on measured temperature and system age. \cite{2021AJ....162..296X} reports a minimum mass of 0.84\,M$_{\rm Jup}$.
      \vspace{2mm}
\end{table*}

\section{Data Reduction \& Analysis}\label{DataRed}

\subsection{Data Reduction}\label{reduce}

The MIRI imaging reduced automatically by the JWST pipeline and available in MAST suffers from a non-uniform background structure at the longest wavelengths. To address this, we reprocessed the data using a custom software package, \texttt{MEOW}, that is available on GitHub\footnote{\url{https://github.com/kevin218/MEOW}. Version 1.0 of \texttt{MEOW} was used for this reduction.  We used JWST pipeline version 1.15.1 and CRDS jwst$\_$1225.pmap.} The background subtraction code is based on a STScI JWebbinar demo\footnote{\url{https://github.com/spacetelescope/jwebbinar_prep/blob/jwebbinar31/jwebbinar31/miri/Pipeline_demo_subtract_imager_background-platform.ipynb}} that produces flat-fielded Stage 2 data with a custom background subtraction using the multiple dithers on each source.

\subsection{White dwarf model}\label{WDmodel}

Determining the planet's contribution to the measured mid-infrared flux requires a precise spectral energy distribution (SED) model of the white dwarf. Comparing the modeled flux to the observed combined star and planet flux allows us to isolate and measure the planet's emission. Thus, before any data analysis can proceed, an accurate model of the white dwarf is essential.

The white dwarf WD~1856 (parameters given in Table \ref{parameters}) is characterized by a temperature of T$_{\rm eff}$ = 4920\,K and classified as a DA spectral type given the presence of a hydrogen line in its spectrum \citep{2021AJ....162..296X}. 
We leverage out-of-transit JWST NIRSpec PRISM observations of WD~1856 from GO~\#2358 (MacDonald et al. submitted) to compute the white dwarf's physical parameters. Considering only the region of the spectrum from $\lambda = 0.6-4.5$\,µm, we minimized the $\chi^2$ for a white dwarf model atmosphere \citep{blouin2018a,blouin2018b} parameterized by the white dwarf's effective temperature, solid angle and hydrogen-to-helium abundance ratio. Combined with the known distance of WD 1856, the solid angle directly constrains its radius, which in turn constrains its mass and surface gravity given white dwarf structure models \citep{2020ApJ...901...93B}. The best-fit solution has a hydrogen-to-helium ratio of 4.1 and yields an H$\alpha$ line that extends 2\% below the continuum, which is consistent with existing optical spectroscopy \citep[][not considered here in our fit]{2021A&A...649A.131A,2021AJ....162..296X}. Assuming a pure hydrogen composition yields a significantly worse fit to the PRISM data. { The median discrepancy between the model and fit, binned to 0.7\,$\mu$m bandwidths (equivalent to the smallest bandwidth used in MIRI imaging), is 1.05\%. To test the model's ability to predict flux at longer wavelengths beyond the fitting region, we repeated the fit using only the $<2.5$\,µm PRISM data. This resulted in a median change of just 0.05\% in the 2.5–5.0\,µm range and 0.7\% in the 5–18\,µm range. We therefore conclude that the model reliably predicts the mid-infrared flux of the white dwarf to about 1\% precision.
}

\begin{table*}
\centering
\caption{ MIRI bands and wavelengths (left two columns), modeled white dwarf flux (third column), measured blended planet plus white dwarf flux with MIRI photometry (fourth column), measured excess or planet flux given by the difference between the model and measured values (fifth column), and percent excess, which is measured flux of the planet divided by the model flux of the white dwarf (last column).}
\begin{tabular}{lcccc|cc}
\hline
{MIRI} & Wavelength & {WD Model} & {Measured} & \multicolumn{2}{c}{Excess (Planet) Flux} \\
Band & (µm) & (µJy) & (µJy) & (µJy) & F$_{\rm p}$/F$_{\rm *}$ (\%) \\
\hline
F560W   & 5.6 & 115.17 & 115.94$\pm$2.35 & 0.77$\pm$2.35 & 0.7$\pm$2.0 \\
F770W   & 7.7 & 66.97 & 68.72$\pm$1.44 & 1.75$\pm$1.44 & 2.6$\pm$2.1 \\
F1000W  & 10.0 & 40.60 & 40.66$\pm$0.87 & 0.06$\pm$0.87 & 0.2$\pm$2.1 \\
F1130W  & 11.3 & 31.67 & 31.69$\pm$0.71 & 0.02$\pm$0.71 & 0.1$\pm$2.2 \\
F1280W  & 12.8 & 25.03 & 25.85$\pm$0.56 & 0.82$\pm$0.56 & 3.3$\pm$2.2 \\
F1500W  & 15.0 & 18.27 & 20.28$\pm$0.57 & 2.01$\pm$0.57 & 11.0$\pm$2.8 \\
F1800W  & 18.0 & 12.87 & 15.18$\pm$0.70 & 2.31$\pm$0.70 & 17.9$\pm$4.6 \\
\hline
\end{tabular}
\label{photometry}
\end{table*}

The cooling age of the white dwarf was determined by evolving a \texttt{MESA} (r23.05.1) model of the appropriate mass down to $T_{\rm eff}=4920\,$K \citep{2023ApJ...950..115B}. A relatively thin hydrogen layer of $\log M_{\rm H}/M_{\star}=-6$ was assumed, consistent with the measured photospheric hydrogen-to-helium ratio \citep{2018ApJ...857...56R}. This results in a white dwarf cooling age of 5.4$\pm$0.7\,Gyr, where the uncertainty includes both the propagated error from the atmospheric parameters and systematic cooling model uncertainties \citep{2024arXiv241014014P}. Using \texttt{wdwarfdate} \citep{2022AJ....164...62K} to estimate the pre-white dwarf lifetime, we find a total age of 9.3$^{+2.8}_{-1.9}$\,Gyr. However, we truncate the upper tail of the distribution to constrain the age to $7.4-10.0$\,Gyr, since the discovery paper \citep{Vanderburg_2020} constrained the age of the system to $<10$\,Gyr using other means. The progenitor of WD~1856 was likely a F or A-type star.

We used our white dwarf model to compute the expected flux from the white dwarf in the MIRI spectral bands. All relevant spectral lines and molecular bands up to 30\,µm are included \citep{Limbach22}. Using the white dwarf model, and the MIRI throughputs and parameters from the JWST \texttt{Pandeia} engine, coupled with the Pandeia RefData version 4.0, we computed the MIRI band fluxes for the white dwarf in all filters where observations were conducted. The modeled flux values in each band are listed in Table~\ref{photometry}.

\subsection{Analysis}

\begin{figure}[b]
\centering
\includegraphics[width=0.46\textwidth]{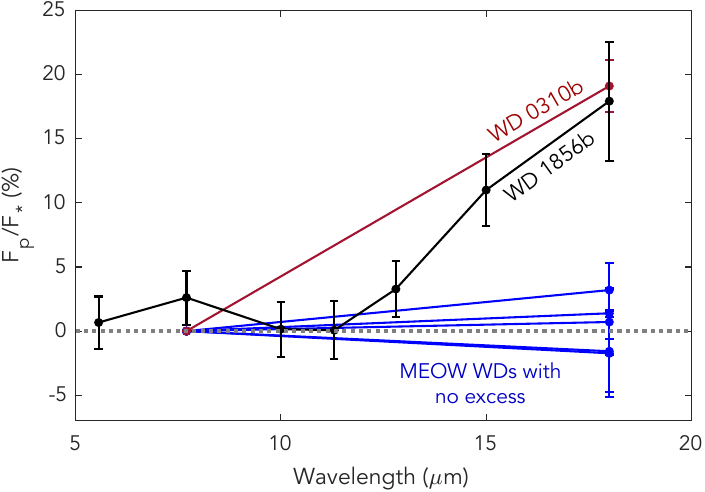}
\caption{Infrared excess of WD~1856 compared with excesses observed in white dwarfs from the JWST MIRI MEOW survey, including the planet candidate around WD~0310-688 \citep{2024ApJ...973L..11L}, which shows a similar excess at 18\,µm to that of WD~1856\,b. Note that the WD~1856 model is anchored to the measured NIRSpec spectrum. In the absence of a NIRSpec spectrum for the MEOW data, we instead anchor the data at 7.7 microns.
}
\label{IRexcessMEOW}
\vspace{3mm}
\end{figure}

Using the reprocessed Stage 2 data from Section \ref{reduce}, we then conducted aperture photometry on the blended white dwarf plus planet point spread function (PSF). To conduct aperture photometry we used the Python \texttt{photutils} package and the aperture correction values provided in the JWST CRDS\footnote{\url{https://jwst-crds.stsci.edu/}; file version \path{jwst_miri_apcorr_0014.fits}}. For the aperture sizes, we used the values corresponding to the full array (as we read out the full MIRI subarray). We replaced \texttt{NAN} values in the stage 2 data using nearest neighbor interpolation. We computed fluxes at each of the four dithers using three different encircled energy aperture sizes: 60\%, 70\%, and 80\% of the white dwarf's PSF. Larger encircled energy apertures are not available for aperture photometry, and smaller apertures are avoided since the measurements are photon-noise limited (rather than systematic-limited) at longer wavelengths. The median flux values, derived from all dithers and encircled energy apertures, are reported in the fourth column of Table~\ref{photometry}.

\begin{figure*}[]
\centering
\includegraphics[width=0.96\textwidth]{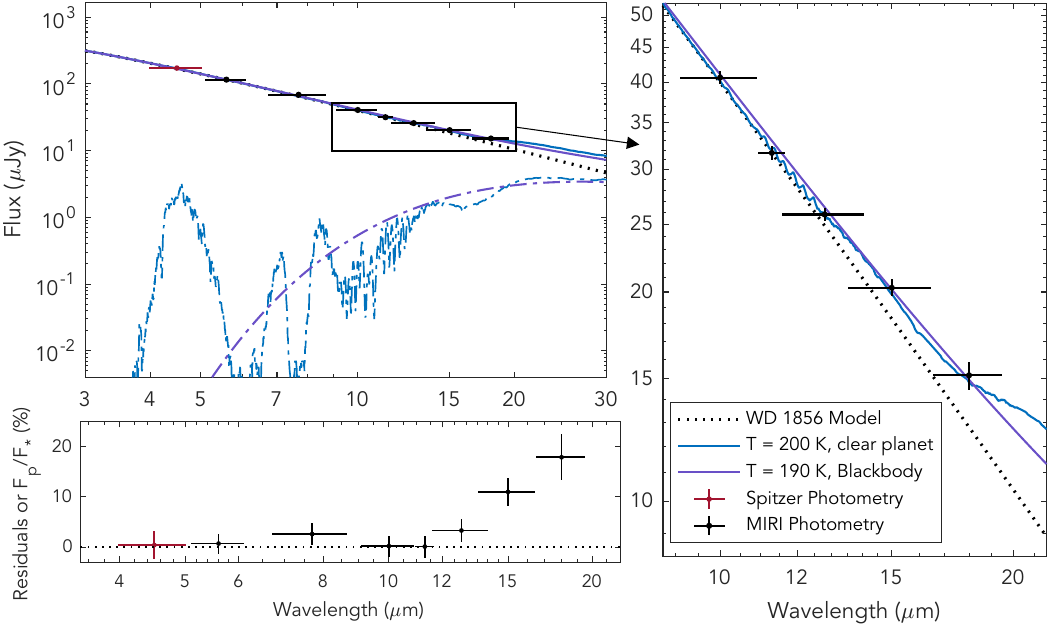}
\caption{{\it Top Left:} Model SED of the white dwarf (dotted black line), a 190\,K blackbody (dash-dotted purple line), and a 200\,K cloudless planet model\citep{Marley2021} (dash-dotted blue line; the coldest model available); the legend applies to all three panels. Lines for the blended SEDs (white dwarf + planet model) are shown in the same color as the planet-only models, but as a solid lines. Photometry from both archival Spitzer and the new MIRI observations are shown as red and black errorbars, respectively. The errorbars in the y-axis are the 1$\sigma$ uncertainty and the errorbars in the x-axis represent the width of the photometric bandpasses. {\it Right:} A zoom-in on the SED in the region where an excess is detected. The MIRI photometry is inconsistent with the WD-only model, and is consistent with WD + planet models. {\it Bottom Left:} Normalized residuals (\%) between the measured photometry and modeled white dwarf photometry show a significant mid-infrared excess at the longer wavelengths due to the exoplanet.}
\label{SED}
\end{figure*}

The measurement error is calculated as the root-mean-square of the standard deviation across all measurements (from all dithers and encircled energy apertures) and a systematic error due to the absolute photometric calibration of the JWST MIRI imager, which we take to be 2\% of the measured flux value \citep{2022AJ....163..267G,2024arXiv240910443G}. We note that the 2\% absolute photometric calibration error quantifies the discrepancy between the measured and modeled white dwarf flux. By definition, this error encompasses the uncertainty associated with the model itself; therefore, a separate error for the model value is not provided in Table~\ref{photometry}. Adopting 2\% as the systematic noise floor also aligns with recent publications on mid-infrared MIRI photometry, which have shown that field white dwarfs without excess match models at the 2--3\% level when measurements have sufficiently high signal-to-noise ratios \citep{2024arXiv240113153M,2024AJ....167..257P}. To illustrate this, we present the infrared excess of WD~1856\,b in Figure \ref{IRexcessMEOW}, alongside the infrared excess observed in white dwarfs from the JWST MIRI Exoplanets Orbiting White Dwarfs (MEOW) survey \citep{2023jwst.prop.4403M}. For the MEOW survey data, the white dwarf models are anchored to the measured flux at 7.7\,µm (as there is no NIRSpec data to anchor the model to, as is the case for WD~1856), with the excess relative to the fitted model shown at 18\,µm. Only white dwarfs from the MEOW sample without strong magnetic fields, extremely bright companion stars, and with 18\,µm measurement precisions $<$ 5\% are included in this comparison. Notably, the planet candidate around WD~0310-688 \citep{2024ApJ...973L..11L} exhibits a comparable level of excess at 18\,µm as WD~1856\,b.

By subtracting the measured flux of WD~1856 from the model white dwarf fluxes we determined the flux excess in each imaging band, representing the planet's flux contribution. We present these values in Table~\ref{photometry}. The excess is reported both as an absolute flux value in microJansky (e.g., measured flux minus the modeled white dwarf flux) and as a percentage flux excess (F$_{\rm p}$/F$_{\rm *}$). No significant flux from the planet is detected at 11.3\,µm or shorter wavelengths. For the longer wavelength observations, the detections have the following statistical significances: 1.5$\sigma$, 3.9$\sigma$ and 3.9$\sigma$ at 12.8\,µm, 15\,µm and 18\,µm, respectively. Combining the results from these imaging bands, we determine that flux from the exoplanet WD~1856\,b is detected with an overall statistical significance of 5.7$\sigma$.

Figure \ref{SED} shows the JWST MIRI and \textit{Spitzer} \citep{Vanderburg_2020} photometry overlaid with blended models of the white dwarf and planet SEDs. The planetary flux excess is evident in the right and bottom left plots. The MIRI photometry matches well with the SED of the white dwarf blended with either a $\sim$190\,K blackbody or giant planet model \citep{Marley2021}.

{ 

We further note that \cite{2024ApJ...973L..11L} found that adjusting the white dwarf model can shift the flux up or down across all mid-IR bands equally but cannot introduce a slope in the mid-IR. Consequently, the slope observed at wavelengths longer than 12.8\,µm cannot be explained by modifications to our white dwarf model. Furthermore, any adjustments to the model would result in a uniform offset in the excess flux at shorter wavelengths (5–12\,µm). A cold planet cannot produce a constant excess across such a broad temperature range, suggesting that our white dwarf model is accurate within the small deviations observed at shorter wavelengths. Specifically, the model predicts flux within 1\% precision, as deviations remain, on average, $<$1\% from measured values in the 5–12\,µm region. This model uncertainty is already accounted for in the 2\% systematic error, as previously mentioned. We note that some cool white dwarfs exhibit poorly understood infrared spectral features \citep{2024arXiv240916224B}. However, if such features were present in WD~1856 we would anticipate as deviations in the NIRSpec data, which are not observed. Therefore, we have no reason to believe that the spectrum of this white dwarf deviates from a Rayleigh-Jeans profile predicted by our model in the mid-infrared.

{\it Could the excess be caused by a source other than the known transiting planet around WD~1856?} Infrared excess around white dwarfs can originate from debris disks, companions (such as low-mass stars, brown dwarfs, or planets), or a spatially unresolved background source. We dismiss the disk scenario as extremely unlikely for two reasons: (1) since the planet is near its equilibrium temperature, a disk at a similar temperature would need to be at a comparable separation, and (2) we observe no evidence of pollution in the atmosphere of the white dwarf, whereas a nearby disk would be expected to scatter debris onto the star, especially given it's proximity to the planet. A PSF subtraction of the star using the same method presented in \cite{2024ApJ...973L..11L} reveals no sources close-in, constrain the excess to be with 0.2" of the white dwarf. This constrains the false positive probability of the source being a background object to $<$0.01\%, which we consider negligible. There are no signs of additional unresolved planets in the system in TTVs \cite{2023MNRAS.521.4679K}, so we conclude that the measured excess must almost certainly be due to the known transiting planet.
}

\section{Results} \label{sec:results}

\subsection{Blackbody Fit to the IR Excess}

To accurately determine the planet’s brightness temperature measured by MIRI in the mid-infrared, we fit a blackbody model to the observed planetary infrared excess. For the blackbody fit we incorporate prior knowledge constraining the planet's radius. The planet is the only known transiting planet around a white dwarf, discovered using NASA's TESS mission, and one of only about five confirmed white dwarf exoplanets\citep{veras2021planetary}. It has an 8-minute-long, 56\% deep grazing transit signal that recurs every 1.41 days\citep{2023MNRAS.521.4679K}. Transit modeling indicates a planet-to-star radius ratio of 7.86$\pm$0.01\citep{2021AJ....162..296X}, which, combined with our white dwarf model, yields a planet radius of 0.946$\pm$0.017\,R$_{\rm Jup}$. We use a Gaussian prior for the planet’s radius, and a uniform prior on the temperature over the range of [100, 600]\,K.  The fit constrains the mid-infrared brightness temperature to 186$^{+6}_{-7}$\,K. The parameters for the white dwarf and exoplanet, as used and determined in this study, are listed in Table \ref{parameters}.

\subsection{Consistency between Planet and Blackbody Models}

\begin{figure*}[h!]
\centering
\includegraphics[width=0.96\textwidth]{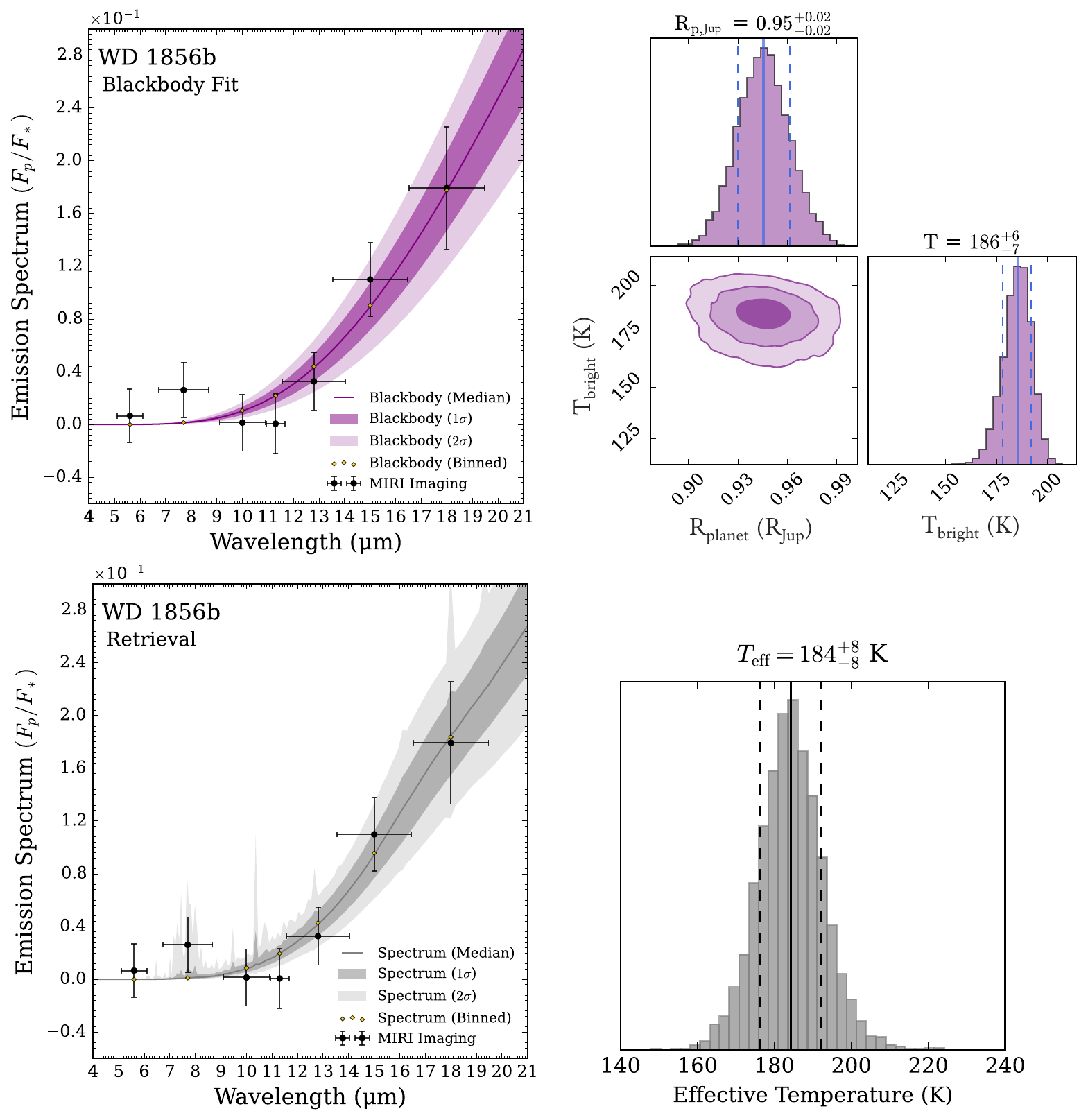}
\caption{{\it Top:} Blackbody fit to the measured planet-star flux ratio (top left) and the corresponding posterior distribution (top right).  {\it Bottom:} Atmospheric retrieval with \texttt{POSEIDON} on the measured photometry (bottom left) and the corresponding computed effective temperature posterior (bottom right). Both approaches constrain the planet’s brightness/effective temperature to $\approx$ 185\,K.
}
\label{posterior}
\vspace{1mm}
\end{figure*}

Because the emission is detected at a significance of only 5.7$\sigma$, we used a simple blackbody model to fit the data. The resulting blackbody fit and the corresponding posterior distribution are shown in the top panels of Figure~\ref{posterior}. However, since planetary spectra can deviate significantly from blackbodies { (sometimes by orders of magnitude, particularly in the 4–5 micron region) we also consider a simple atmospheric model. This fit demonstrates that the large deviations seen in the 4–5 micron region are not present in the mid-IR and that the derived effective temperature from the atmospheric model remains consistent with the brightness temperature inferred from the blackbody model.}

We conduct an atmospheric retrieval on the MIRI photometric data with the \texttt{POSEIDON} package \citep{2017MNRAS.469.1979M,2023JOSS....8.4873M}. Given the limited number of data points (7), with only 2 data points containing significant emission, we construct a simple atmospheric model. We assume a H$_2$-He dominated atmosphere (with a He/H$_2$ number density ratio of 0.17) and parameterize the abundances of NH$_3$ and CH$_4$ (given their prominence in mid-IR spectra of Jupiter; \citealt{2016Icar..278..128F}). Our atmospheric models range from 10$^{-7}$--100\,bar, with 100 layers spaced linearly in log-pressure, with a vertical temperature gradient for the pressure-temperature profile. The temperature profile is defined by the temperature at two reference pressures (a `top' pressure at 10$^{-5}$\,bar and a `base' pressure at 10\,bar), with a linear in log-pressure temperature gradient connecting the two pressures and an isothermal region above and below the reference pressures. We also fit for the planetary reference radius at 10\,bar, adopting the same priors on radius and temperature as used for the blackbody fit. We explored the 5-parameter space using \texttt{PyMultiNest} \citep{2009MNRAS.398.1601F} with 2000 live points.

Figure~\ref{posterior} (lower panels) shows the results of the atmospheric retrieval. As expected, given the low statistical significance of the observations, this more complex model is statistically indistinguishable from the blackbody-only model ($\Delta \ln{\mathcal{Z}} = 0.4$ in favor of the atmosphere model, where $\mathcal{Z}$ is the Bayesian evidence). Nevertheless, we used the posterior samples of the atmospheric properties to estimate the planet's effective temperature. To accomplish this, we extrapolated model spectra corresponding to each set of atmospheric properties over the wavelength range $\lambda = 0.2-250$\,µm and computed the integrated flux for each model. From the computed bolometric flux of each sampled atmosphere, we determined the corresponding effective temperature and subsequently derived the distribution of effective temperatures across all the samples. We thus find an effective temperature for WD~1856\,b of $184\pm8$\,K, consistent with the temperature obtained from the blackbody fit. Future spectroscopic observations with JWST could provide data with sufficient quality for a more meaningful atmospheric retrieval in the mid-infrared.

\subsection{Planet Mass}

Using the fitted brightness temperature of 186$^{+6}_{-7}$\,K, along with the total system age of $7.4-10$\,Gyr and planetary evolution models \citep{2007ApJ...659.1661F,Marley2021}, we estimate the planet’s mass to be 5.2$^{+0.7}_{-0.8}$\,M$_{\rm Jup}$ assuming insolation contributes negligibly to its thermal evolution. { Because the system is older, planetary cooling occurs slowly, and the 1-sigma uncertainty in age corresponds to a $\lesssim$20\% uncertainty in mass.} If the insolation contribution is not negligible, this mass value represents the upper limit of the allowed mass. In that case, using the minimum mass value of 0.84\,M$_{\rm Jup}$ reported in \cite{2021AJ....162..296X}, the mass of WD~1856\,b is constrained to 0.84-5.9\,M$_{\rm Jup}$. In either case, the mass is well below the deuterium-burning limit of 13\,M$_{\rm Jup}$ \citep{2011ApJ...727...57S}, and thus firmly establishes WD~1856\,b's classification as an exoplanet. Previously \textit{Spitzer} observations \citep{Vanderburg_2020} yielded a constraint on the planet's flux at 4.5\,µm, finding the white dwarf's flux to be 173$\pm$10\,µJy and the planet's flux to be 0.7$\pm$4.9\,µJy, which set a 1$\sigma$ upper limit on the planet’s temperature at 250\,K. In the original discovery paper, the tail of the planetary temperature distribution did not fully rule out masses within the low-mass brown dwarf range, leaving the planetary nature of the companion unconfirmed prior to this work.

\section{Discussion \& Conclusions} \label{sec:discuss}

{ In this paper we:
\begin{itemize}
    \item Reported the detection of thermal emission from WD 1856+534~b consistent with a temperature of $186^{+6}_{-7}$ K, making it the coldest exoplanet from which light has been directly detected.
    \item Confirm the planetary nature of WD 1856+534~b: based on the measured temperature and age of the system (9.3$^{+0.7}_{-1.9}$\,Gyr), we constrain the planets mass to 5.2$^{+0.7}_{-0.8}$\,M$_{\rm Jup}$ assuming insolation contributes negligibly to its thermal evolution, or $0.84-5.9$\,M$_{\rm Jup}$ if insolation is a contributing factor. Our estimate of mass takes into account the uncertainties in the age of the system and the temperature of the planet. It is now the first intact exoplanet confirmed within a white dwarf’s “forbidden zone.” 
\end{itemize}
}

\begin{figure}
\centering
\includegraphics[width=0.47\textwidth]{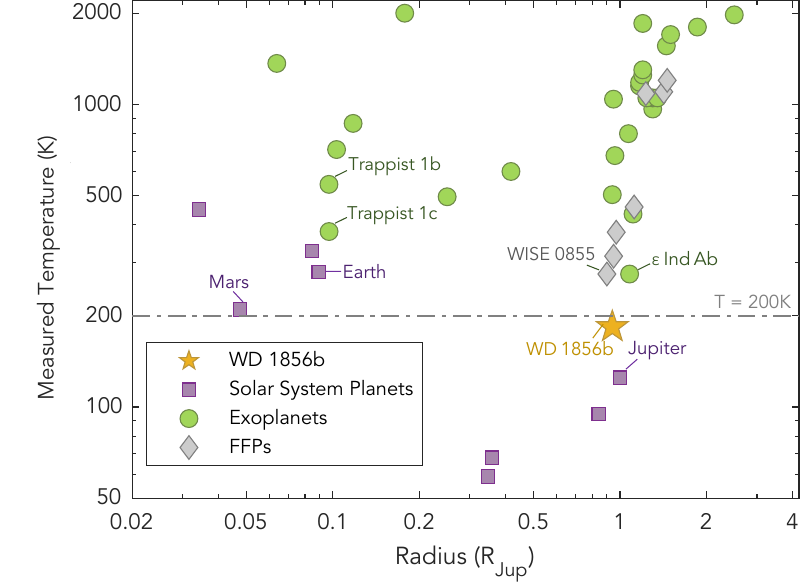}
\caption{Planet radius (in Jupiter radii) versus measured temperature of confirmed exoplanets (green circles) and free-floating planetary-mass objects (FFPs; gray diamonds), solar system planets (purple squares) and WD~1856\,b (yellow star). WD~1856\,b bridges the temperature gap between Jupiter and exoplanets/FFPs with temperature measurements.}
\label{tempVrad}
\end{figure}

Our confirmation of WD~1856\,b as an exoplanet and measurement of its temperature identifies it as the coldest known exoplanet with directly detected emission. Compared to the previous record-holder, $\epsilon$ Ind Ab (\citealt{2024Natur.633..789M}; T = 275\,K, M = 6.3\,M$_{\rm Jup}$, age = 3.5\,Gyr), WD~1856\,b is colder, older, and less massive, demonstrating JWST's capability to detect exoplanets with properties similar to the giant planets in our own Solar System. Moreover, WD~1856\,b is nearly 100\,K colder than the coldest known brown dwarf, or free-floating planetary-mass object, WISE 0855-0714 \citep{2024AJ....167....5L}, which is about the same temperature as $\epsilon$ Ind Ab (see Figure \ref{tempVrad}\footnote{Measured exoplanet and FFP data are from the Exoplanet Archive and \url{exoplanet.eu} on December 8, 2024.}). WD~1856\,b is only about 60\,K warmer than Jupiter's effective temperature of 124.4$\pm$0.3\,K \citep{2023RemS...15.1811R} and about 25\,K colder than Mars \citep{HaberleBook}, which positions WD~1856\,b as a bridge between the previously characterized warmer exoplanets and the giant planets of our Solar System.

WD~1856\,b is the first intact exoplanet confirmed to orbit within the ``forbidden zone" of a white dwarf—the region that would have been engulfed by the progenitor star during its red giant phase \citep{2013MNRAS.432..500N}. When the host star exhausted its nuclear fuel and evolved into a red giant, it expanded far beyond WD~1856\,b's current orbital radius, and must have engulfed any planets in its path. Therefore, WD~1856\,b must have originally orbited far from its host star and migrated close to the white dwarf after the red giant phase. 
Several theories attempt to explain how this migration occurred. One possibility is that WD~1856\,b underwent common-envelope evolution \citep{2021MNRAS.501..676L,2021MNRAS.502L.110C}, spiraling inwards through the star's outer layers before ejecting them and settling into its current orbit. Alternatively, gravitational interactions with other planets \citep{2021MNRAS.501L..43M} or nearby stars \citep{2020ApJ...904L...3M,2021MNRAS.501..507O,2021ApJ...922....4S} after the white dwarf formed could have perturbed WD~1856\,b into a high-eccentricity orbit, which then tidally circularized.
While the common-envelope migration mechanism is feasible for brown dwarfs, it becomes challenging in the planetary mass regime. Although the measured mass from this observation does not rule out common envelope evolution \citep{2021MNRAS.501..676L}, future follow-up observations to characterize the planet's atmosphere could significantly enhance our understanding of its dynamical and migration history.

Looking ahead, the JWST MIRI program that produced these WD~1856\,b observations also includes a second epoch of observations scheduled for July 2025. This phase will be essential for identifying additional companions through common proper motion, which could help reveal whether another planet perturbed WD~1856\,b into its current orbit via high-eccentricity migration. 
Upcoming results from Cycle 1 NIRSpec observations will provide an initial characterization of the planet's atmosphere \citep{2021jwst.prop.2507V,2021jwst.prop.2358M}. Further atmospheric characterization of WD~1856\,b, through spectrally resolved emission measurements, transmission spectroscopy, or rotation curve monitoring, would allow us to explore atmospheric physics in this unprecedented temperature regime. Given its colder temperature, we expect WD~1856\,b's atmosphere to contain different abundances of oxygen, carbon, and nitrogen bearing molecules compared to the warmer gas giant exoplanets characterized to date.
Because the white dwarf host star emits minimal emission in the ultra-violet, WD~1856\,b will also provide unique insight into the atmosphere of a gas giant with limited photochemistry.

In addition to being very cold, WD~1856\,b is relatively distant at 24.8\,pc, making its apparent brightness very low---much fainter than $\epsilon$ Ind Ab at 3.6\,pc. Nevertheless, JWST was able to detect it confidently in less than an hour of total integration time. For other systems with wider orbit (spatially resolved) planets, JWST MIRI has the potential to to directly image even colder gas giants in closer systems, with temperatures as low as 75\,K \citep{2024jwst.prop.6122B}. This capability could provide an opportunity to explore the atmospheres and thermal properties of planets with temperatures comparable to the coldest gas giants in our Solar System. The detection of WD~1856\,b, with its sub-200\,K temperature, marks significant progress in directly detecting exoplanets with thermal emission similar in temperature to Solar System gas giants.

\section*{Acknowledgement}
{ The JWST data presented in this article were obtained from the Mikulski Archive for Space Telescopes (MAST) at the Space Telescope Science Institute. The specific observations analyzed can be accessed via \dataset[doi: 10.17909/t8t4-6775]{https://doi.org/10.17909/t8t4-6775}.} This research has made use of the SIMBAD database,
operated at CDS, Strasbourg, France \citep{2000A&AS..143....9W}.
This research has made use of the NASA Exoplanet Archive, which is operated by the California Institute of Technology, under contract with the National Aeronautics and Space Administration under the Exoplanet Exploration Program. This research has made use of data obtained from or tools provided by the portal \url{exoplanet.eu} of The Extrasolar
Planets Encyclopaedia.
This work is based on observations made with the NASA/ESA/CSA James Webb Space Telescope. The data were obtained from the Mikulski Archive for Space Telescopes at the Space Telescope Science Institute, which is operated by the Association of Universities for Research in Astronomy, Inc., under NASA contract NAS 5-03127 for JWST.  These observations are associated with program \#5204. S. X. is supported by the international Gemini Observatory, a program of NSF NOIRLab, which is managed by the Association of Universities for Research in Astronomy (AURA) under a cooperative agreement with the U.S. National Science Foundation, on behalf of the Gemini partnership of Argentina, Brazil, Canada, Chile, the Republic of Korea, and the United States of America.

\facilities{\it JWST}.

\software{\texttt{MULTINEST} \citep{2009MNRAS.398.1601F}, {\tt astro.py} \citep{2013A&A...558A..33A, 2018AJ....156..123A, 2022ApJ...935..167A}, {\tt numpy.py} \citep{5725236}, {\tt wdwarfdate} \citep{2022AJ....164...62K}, {\tt POSEIDON} \citep{2017MNRAS.469.1979M,2023JOSS....8.4873M}, {\tt SAOImage DS9} \citep{2000ascl.soft03002S}, \texttt{MESA} (r23.05.1)\citep{2023ApJ...950..115B}}

\appendix
\vspace{-6mm}

\section{MIRI Image}\label{AppB}
\begin{figure*}[h]
\centering
\includegraphics[width=1.0\textwidth]{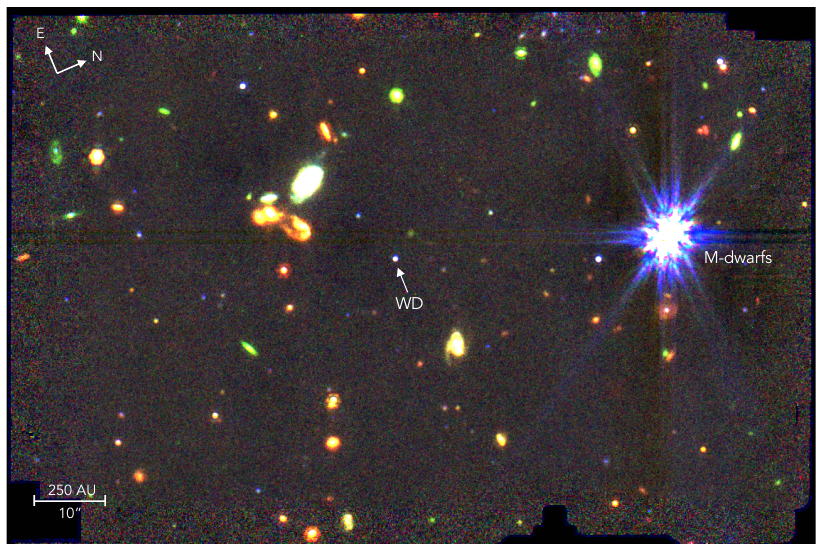}
\caption{False-color image of the WD~1856+534 system constructed using three JWST/MIRI imaging bands: F770W (7.7\,µm, blue), F1500W (15\,µm, green), and F1800W (18\,µm, red). The FOV measures 74$'' \times$ 113$''$, corresponding to 1835\,AU $\times$ 2802\,AU at the system's distance of 24.8\,pc. The (blended) white dwarf and planet are the bluish-white star at the center of the image. The two bright stars near the right-hand side of the field of view (FOV) are mid-M dwarf companion stars to the white dwarf, at a projected separation of 1030\,AU.  Numerous colorful background stars and galaxies are visible throughout the image.}
\label{MIRIim}
\vspace{3mm}
\end{figure*}

A false-color image, created from the MIRI data in three bands ($\lambda$ = 7.7, 15, and 18 µm), is shown in Figure \ref{MIRIim}.

\section{A Note on the Planet's Orbital Phase}\label{AppA}

WD~1856\,b is close enough to the white dwarf for us to expect it to be tidally synchronized, with constant day-night-sides, similar to a hot Jupiter. Since this observation was conducted at a specific orbital phase, it is important to consider whether the phase might influence the measured temperature. For a hot Jupiter, over the orbital phase range 0.32-0.35 we would be observing more of the dayside and so a generally hotter region of the planet. Since WD~1856\,b has a much cooler atmosphere, however, its radiative timescales will be much longer than for hot Jupiters, on the order of 20 days (assuming T=184\,K, $g=140$ m/s$^2$, a hydrogen/helium-dominated atmosphere, and evaluated at $P=0.25$ bar; \citealt{Showman2002}). A simple extrapolation from the hot Jupiter regime might then predict this planet to have uniform temperatures around the globe \citep{Komacek2016}, but those trends assume that cooler planets will be also rotating more slowly. 
Given its fast rotation rate, relatively high gravity, and cool temperatures, we can use the analytic scalings derived for synchronously rotating hot Jupiters \citep{PerezBecker2013,Komacek2016} to compare the planet's radiative timescale to its rotational and wave adjustment timescales. The wave adjustment timescale, $\sim$8 days, characterizes how quickly atmospheric dynamics can reduce temperature gradients. We find that this planet lies near the boundary between regimes where significant day-night temperature differences might or might not exist.
It is therefore ambiguous whether or not the dayside of this planet should be hotter and brighter than the nightside, especially since we do not know how well the assumptions used in the hot Jupiter scale apply here (e.g., the use of a Kelvin wave speed for the wave adjustment timescale). If we are seeing a somewhat brighter dayside, then that may explain why the measured effective temperature is larger than the equilibrium value. In contrast, if the planet's temperature is fairly uniform, then measuring an effective temperature larger than equilibrium implies some significant heating from the planet interior, which may also change the nature of how the atmospheric circulation is driven. It will be valuable to gain more data on this planet, especially at other orbital phases, and pursue more detailed modeling.

\bibliography{main}{}
\bibliographystyle{aasjournal}
\end{document}